\documentclass{iau}
\usepackage{graphicx}

\title[Dwarf spheroidal galaxies] 
{Dark matter distribution and dynamics\\
of dwarf spheroidal galaxies}

\author[Ewa L. {\L}okas]   
{Ewa L. {\L}okas$^1$}

\affiliation{$^1$Nicolaus Copernicus Astronomical Center, Polish Academy of Sciences,\\
Bartycka 18, 00-716 Warsaw, Poland \\ email: {\tt lokas@camk.edu.pl} }

\pubyear{2019}
\volume{353}  
\jname{Galactic Dynamics in the Era of Large Surveys}
\editors{M. Valluri, \& J. A. Sellwood  }
\begin{document}

\maketitle

\begin{abstract}
I review the current status of dynamical modelling of dwarf spheroidal galaxies focusing on estimates of their
dark matter content. Starting with the simplest methods using the velocity dispersion profiles I discuss the
inherent issues of mass-anisotropy degeneracy and contamination by unbound stars. I then move on to methods of
increasing complexity, aiming to break the degeneracy, up to recent applications of the Schwarzschild orbit
superposition method. The dynamical modelling is placed in the context of possible scenarios for the formation of
dwarf spheroidal galaxies, including the tidal stirring model and mergers of dwarf galaxies. The two scenarios are
illustrated with examples from simulations: a comparison between the tidal evolution of dwarfs with cuspy and cored
dark matter profiles and the formation of a dwarf spheroidal with prolate rotation.
\end{abstract}

\firstsection 
\section{Introduction}

Dwarf spheroidal (dSph) galaxies are a class of faint galaxies mostly found in the Local Group, in the vicinity of the
Milky Way and Andromeda. They are gas poor, supported mostly by random motions and believed to be strongly
dominated by dark matter. Their unique property is that resolved stellar populations are available for study i.e.
positions and radial velocities of individual stars can be measured. Over the last two decades dSph
galaxies have been the subject of intensive study, with their kinematic samples increasing from few tens to few hundred
and now a few thousand of stars with measured velocities for best-studied dwarfs. This allowed us to develop more and
more advanced methods of modelling their dark matter content. At the same time, a few scenarios for their dynamical
evolution were proposed. In the first part of this contribution I summarize the development of the mass modelling
methods, from the simplest to the most advanced, while the second part is devoted to a short description of the
formation scenarios.

\section{Mass modelling}

The simplest observable that can be constructed for dSph galaxies in order to obtain an estimate of their mass content
is the line-of-sight velocity dispersion of the stars or its profile as a function of galactocentric radius. This
profile can be modelled by solving the Jeans equation which relates the velocity dispersion to the underlying
gravitational potential and the velocity anisotropy parameter $\beta$. The anisotropy parameter measures the amount of
radial versus tangential motion in stellar orbits and can take the values between $\beta=1$ (radial orbits) and
$\beta=-\infty$ (circular orbits) with $\beta=0$ corresponding to isotropic orbits.

Early studies have demonstrated that dSph galaxies are strongly dominated by dark matter so that their mass-to-light
ratios exceed the stellar values by at least one or two orders of magnitude. It has also been shown, e.g. for the
best-studied Fornax dSph galaxy (\cite[{\L}okas 2002]{Lokas2002}), that the velocity dispersion profile can be equally
well reproduced by different combinations of dark matter density profiles and anisotropy parameters. In general,
shallower profiles require more radial stellar orbits and steeper profiles require more tangential orbits to fit a
given data set. This is a manifestation of the so-called mass-anisotropy degeneracy inherent in the modelling of
any pressure-supported systems, from dSph galaxies to galaxy clusters. For a long time, the simplest and customary way
to go around the problem was to assume the stellar orbits to be isotropic in order to infer the mass distribution in
dSph galaxies. Such an assumption would obviously lead to biases in the estimated density profiles if the real
anisotropy was different. Simulations of galaxy formation and evolution have since demonstrated that the anisotropy
indeed tends to depend on radius and departs from zero.

The mass-anisotropy degeneracy can be to some extent broken by adding constraints from a higher velocity moment,
the kurtosis, that could be measured from more numerous samples of stellar kinematics which became available. Modelling
the kurtosis involves solving the higher order Jeans equation. Since kurtosis is mostly sensitive to the anisotropy, in
combination with the modelling of dispersion, both moments can constrain both the mass distribution and anisotropy.
However, this still usually requires some assumptions concerning the functional form of orbital anisotropy, for example
$\beta$ to be constant with radius (\cite[{\L}okas et al. 2005]{Lokas2005}, \cite[{\L}okas 2009]{Lokas2009}).

The second essential issue in the mass modelling of dSph galaxies is the question of stellar membership. In order to
obtain a reliable mass estimate it is important to model a clean sample, i.e. composed only of member stars. If the
sample is contaminated by tidally stripped stars or background and foreground stars from the Milky Way that do not trace
the potential of the dwarf galaxy, the results will be biased. Such biased estimates will typically lead to the
inference of more tangential orbits or more extended mass distributions since unbound stars tend
to increase the velocity dispersion at larger galactocentric radii (\cite[Klimentowski et al. 2007]{Klimentowski2007},
\cite[{\L}okas et al. 2008]{Lokas2008}).

Recently, more advanced methods of mass modelling of dSph galaxies have been developed. In the Schwarzschild orbit
superposition method the velocity moments are reconstructed from a combination of stellar orbits. No assumptions on the
anisotropy are required and the anisotropy profile follows as a result of the application of the method. For spherical
dwarfs the method works very well and is able to reproduce different anisotropy profiles as well as density profiles of
simulated galaxies (\cite[Kowalczyk et al. 2017]{Kowalczyk2017}). For non-spherical objects application of the method
with the assumption of spherical symmetry is subject to an inherent bias depending on the line of sight; in particular,
the anisotropy is underestimated, and more so for the line of sight along the shortest axis of the stellar component
(\cite[Kowalczyk et al. 2018]{Kowalczyk2018}). A recent application of the Schwarzschild method to the modelling of the
Fornax dwarf used around 3300 carefully selected stars from combined catalogues and allowed to constrain both the mass
and anisotropy profiles in the dwarf. The mass-to-light ratio was found to increase with radius in agreement with
previous estimates and the anisotropy profile was found to be slightly decreasing with radius, but consistent with
isotropy at all radii (\cite[Kowalczyk et al. 2019]{Kowalczyk2019}).

Future improvements in the mass modelling of dSph galaxies should include: increasing the samples of member stars to
reduce sampling errors, developing better membership determination methods, using multiple stellar populations,
including proper motion measurements for the stars (\cite[Strigari et al. 2018]{Strigari2018}) and extensions to
non-spherical models (\cite[Hayashi et al. 2018]{Hayashi2018}).

\section{Formation scenarios}

The dynamics of dSph galaxies is best elucidated by studying their formation scenarios. Although these objects could
have formed in isolation, the scenarios most studied in recent years involved interactions with other systems. The two
main scenarios considered were the tidal stirring due to the proximity of the Milky Way and mergers between dwarfs. In
both scenarios the progenitors of dSph galaxies were disky dwarfs. The tidal stirring model (\cite[Mayer et al.
2001]{Mayer2001}) explains the density-morphology relation observed among the Local Group dwarfs, i.e. the fact that
dSph galaxies are situated close to the Milky Way or Andromeda, tend to be stripped of gas, spheroidal and non-rotating
while dwarf irregular galaxies are typically found at larger distances, contain gas, are more flattened and rotate. The
merger scenario was invoked to explain peculiar structural and kinematic features found in some dwarfs.

\subsection{Tidal stirring}

\begin{figure}
\begin{center}
\includegraphics[width=2.5in]{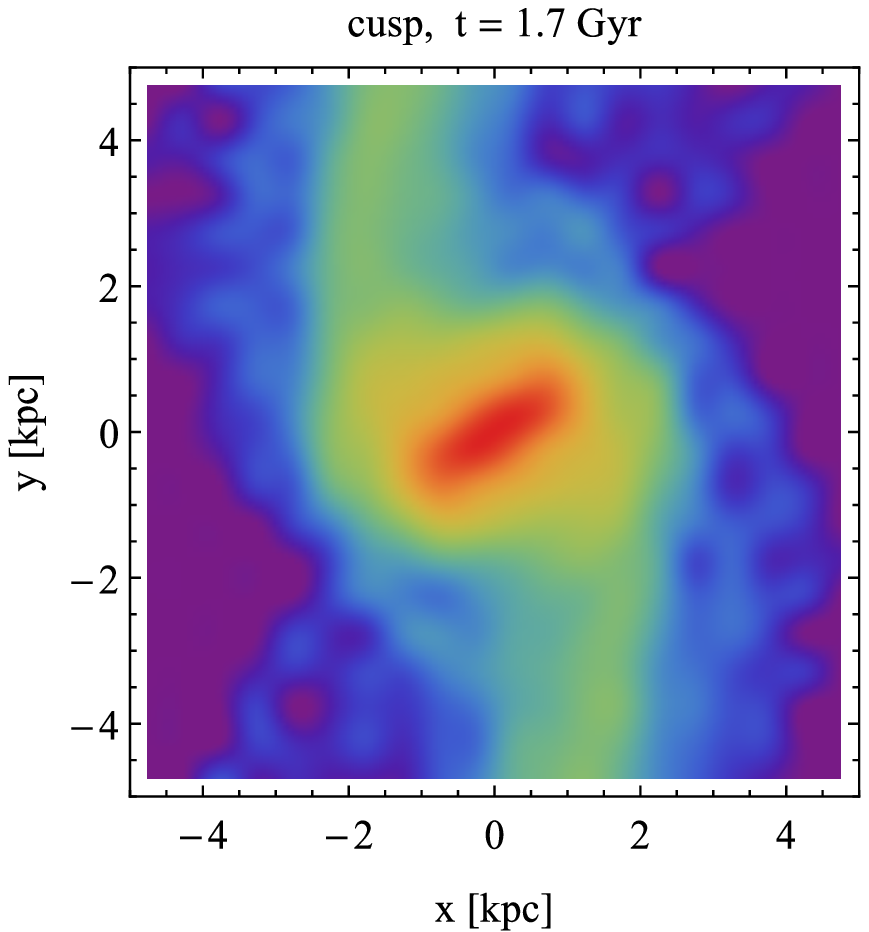}
\includegraphics[width=2.5in]{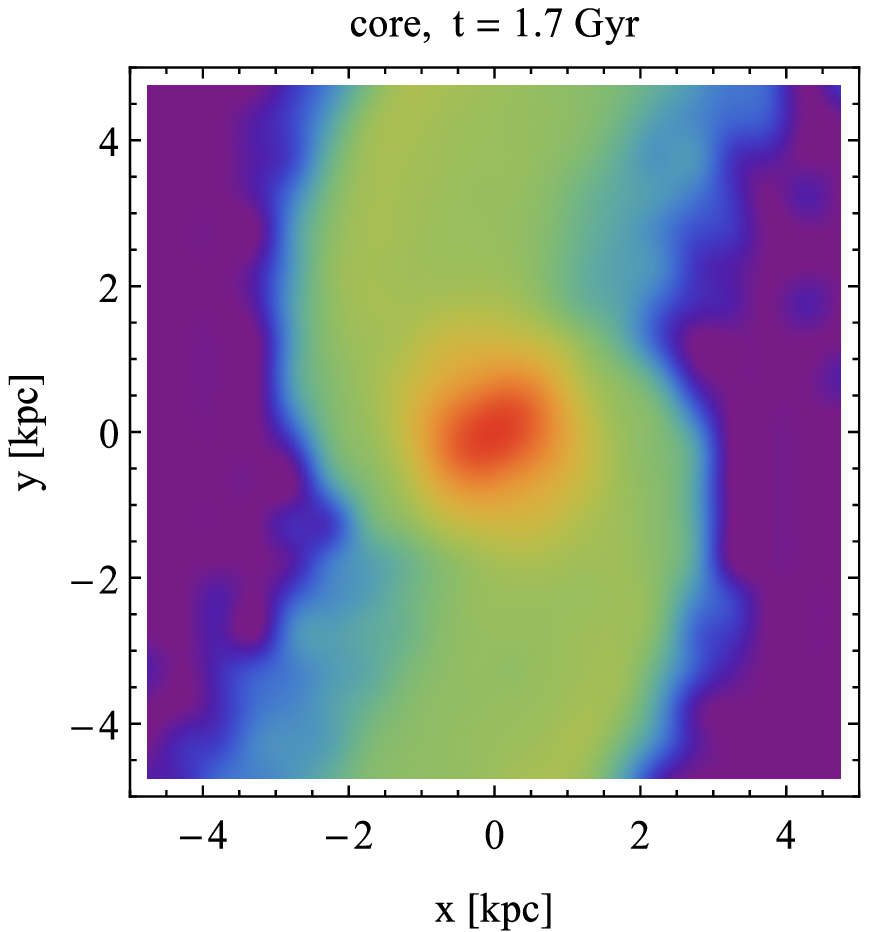} \\
\vspace*{0.5 cm}
\includegraphics[width=2.5in]{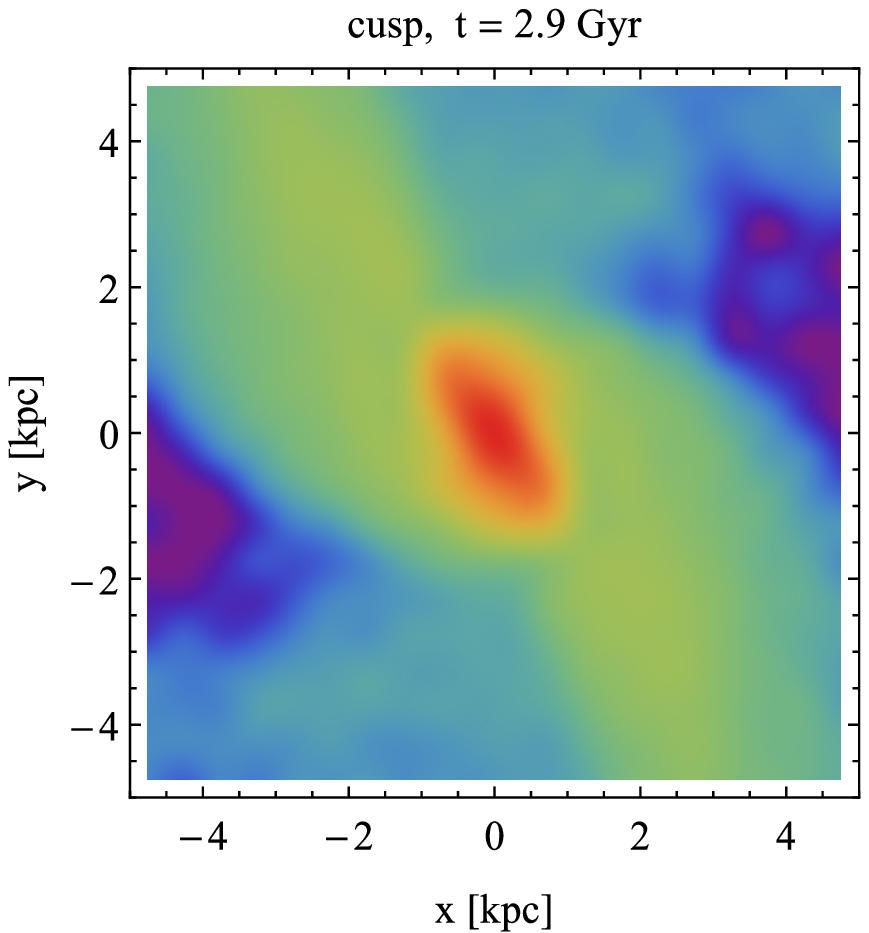}
\includegraphics[width=2.5in]{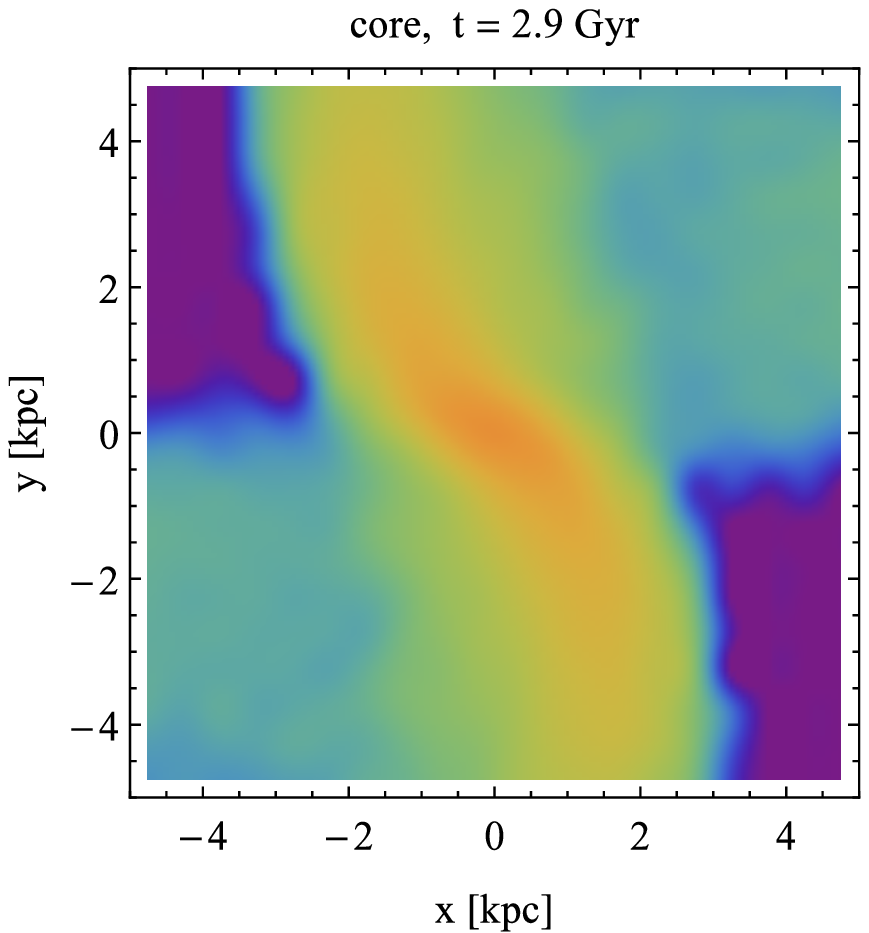}
\caption{Effect of tidal stirring on two initially disky dwarf galaxies evolving around the Milky Way-like galaxy. Both
dwarfs were placed on a prograde orbit with apo- to pericenter ratio equal to 100 kpc/20 kpc. The initial structural
parameters of the dwarfs differed only by their initial dark matter profile. The dwarf with the initial cuspy
($r^{-1}$) density profile (left panels) forms a bar at the first pericenter passage and survives for 10 Gyr. The dwarf
with a cored ($r^{-0.2}$) profile (right panels) dissolves after the second pericenter passage (\cite[{\L}okas
2016]{Lokas2016}). The stellar components of both dwarfs are shown in projection onto the orbital plane after the first
(upper row) and second (lower row) pericenter.}
\label{cuspcore}
\end{center}
\end{figure}

\begin{figure}
\begin{center}
\includegraphics[width=4.5in]{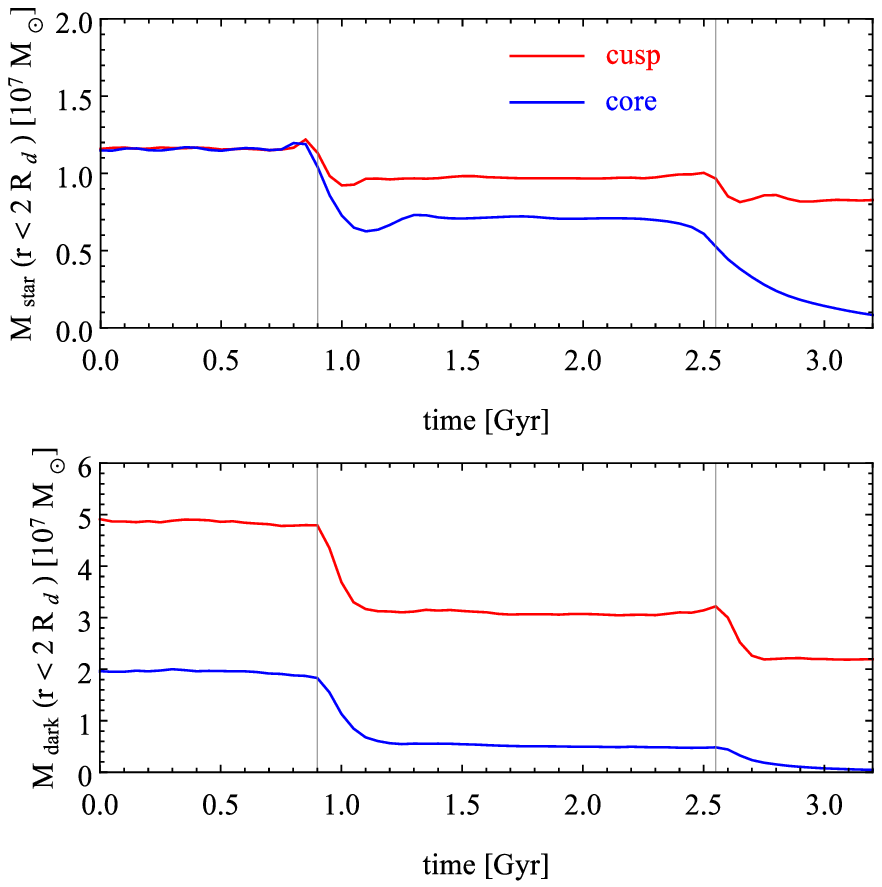}
\caption{Evolution of the mass contained within two disk scale-lengths $2 R_d = 0.82$ kpc for stars (upper panel) and
dark matter (lower panel) for the dwarfs with initially cuspy (red upper line) and cored (blue lower line) dark matter
profile. The vertical gray lines indicate pericenter passages on an orbit with apo- to pericenter of 100 kpc/20 kpc.
For the cored dwarf both masses approach zero after the second pericenter passage signifying the disruption of the
dwarf.}
\label{masses}
\end{center}
\end{figure}

In simulations based on the tidal stirring scenario an initially disky dwarf galaxy is placed on a typical eccentric
orbit around the Milky Way-like galaxy and evolved for a few gigayears. The effects of tidal stirring involve the mass
loss (in both dark matter and stars), the morphological transformation of the stellar component from a disk to a
spheroid and the change of rotation into random motions of the stars. The final product of tidal stirring depends
strongly on the size of the orbit: it can leave the dwarf's disk almost unaffected for an extended orbit or transform it
to a small perfect sphere for a tight orbit (\cite[Kazantzidis et al. 2011a]{Kazantzidis2011a}). The evolution also
critically depends on the initial dark matter distribution in the dwarf. The dwarfs with an initial cuspy dark matter
profile tend to survive for many gigayears while those with a cored profile dissolve after one or two pericenter
passages. A comparison between these two cases is illustrated in Figs~\ref{cuspcore} and \ref{masses}.

An interesting intermediate stage of the evolution from a disk to a spheroid involves the formation of a tidally
induced bar in the dwarf (\cite[{\L}okas et al. 2014a]{Lokas2014a}, \cite[Gajda et al. 2017]{Gajda2017}). The bar
always forms at the first pericenter passage and becomes shorter in time but the spheroidal shape usually survives
until the end of evolution which may explain the non-spherical appearance of dSph galaxies. If the gas is included in
the progenitor dwarf the tidally induced stellar bars are weaker but the effect is strong only in gas-dominated dwarfs
(\cite[Gajda et al. 2018]{Gajda2018}). On the other hand, if the gas was stripped by ram pressure in the hot halo of
the Milky Way then the bars would survive.

Obvious candidates for tidally stirred dwarfs in the vicinity of the Milky Way include the Sagittarius and Carina
dwarfs, although most of dSph galaxies in the Local Group show signs of tidal extensions. In the case of the Sagittarius
dwarf it has been demonstrated that its present structural and kinematic properties can be reproduced by a model in
which the initially disky progenitor has just passed the second pericenter on its orbit around the Milky Way
(\cite[{\L}okas et al. 2010]{Lokas2010}). In this model the stellar component still preserves its bar-like shape
(slightly inclined with respect to the orbit) and shows very little rotation. The recent compilation of data for the
Carina dSph galaxy, showing its elongated shape and significant rotation, can also be reproduced by a late-stage
tidally stirred disky dwarf (\cite[Fabrizio et al. 2016]{Fabrizio2016}).

\subsection{Mergers of dwarfs}

Some dwarfs show features that cannot be explained by tidal stirring, including e.g. stellar shells in Fornax. In
addition, there are distant dSph galaxies in the Local Group that probably did not have enough time to strongly interact
with the Milky Way or Andromeda (e.g. Cetus and Tucana). Besides, in tidally stirred dwarfs the remnant rotation is
always around the short axis of the stellar component while some dwarfs show rotation around the long axis (prolate
rotation), e.g. Andromeda II (\cite[Ho et al. 2012]{Ho2012}) and Phoenix (\cite[Kacharov et al. 2017]{Kacharov2017}).
Such features point toward a different scenario for the formation of at least some dSph galaxies, such as mergers
between initially disky dwarfs (\cite[Kazantzidis et al. 2011b]{Kazantzidis2011b}). This possibility is supported by
constrained simulations of the Local Group, aiming to reproduce the properties of both big galaxies, the Milky Way and
Andromeda, as well as their satellite population. In particular, it has been demonstrated that in such environments a
significant fraction of subhaloes experienced a strong interaction with another subhalo (\cite[Klimentowski et al.
2010]{Klimentowski2010}). Since the typical velocities of single subhaloes around the Milky Way are rather too large
for such interactions to be significant, mergers usually occur between subhaloes that were accreted in pairs, whose
relative velocities are low.

\begin{figure}
\begin{center}
\includegraphics[width=4.6in]{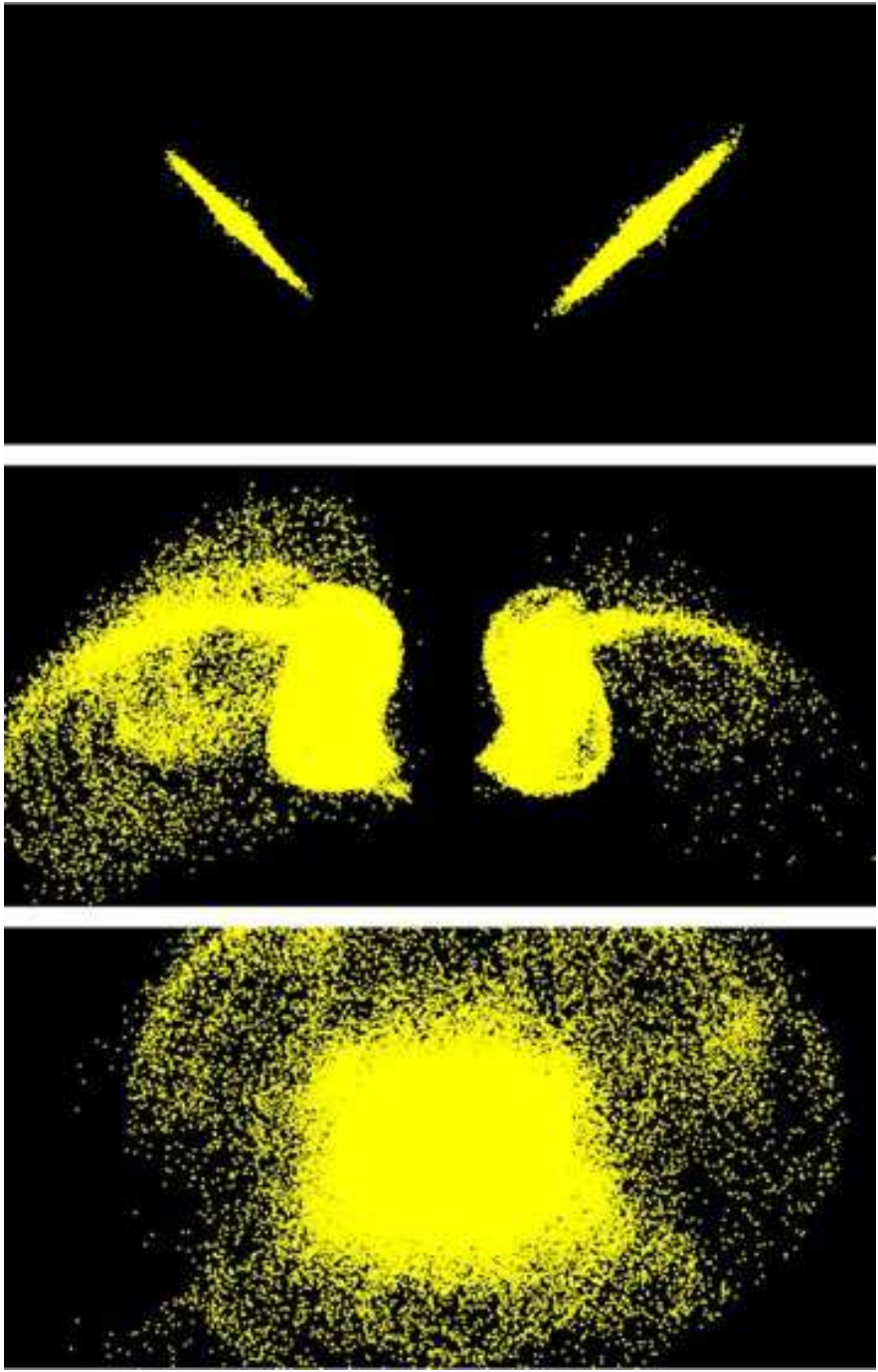}
\vspace*{0.5 cm}
\caption{Formation of a dSph galaxy similar to Andromeda II by a merger of two initially disky dwarfs (\cite[{\L}okas
et al. 2014b]{Lokas2014b}). The three panels show the stellar components in three stages of the merger: before, with
the two galaxies approaching (upper panel), after the first encounter (middle panel) and at the end of evolution, after
the merger (lower panel). The image in each panel has the size of 40 kpc $\times$ 20 kpc.}
\label{merger}
\end{center}
\end{figure}

The merger scenario has been proposed as a way to explain the origin of prolate rotation in Andromeda II
(\cite[{\L}okas et al. 2014b]{Lokas2014b}). In this scenario, the progenitors
are two disky dwarfs of equal mass and similar structural parameters approaching each other on a radial orbit. Their
disks are initially inclined at the right angle but both have significant components of their angular momenta aligned
with the merger axis. These parts of angular momenta are preserved after the merger in the form of prolate rotation. The
remnant dwarf rotates around its long axis while the rotation around the short axis is close to zero. Three different
stages of such a merger are illustrated in Fig.~\ref{merger}. The model reproduces well the non-spherical shape of the
stellar component and if the gas is included (\cite[Fouquet et al. 2017]{Fouquet2017}) also the presence and
distribution of different stellar populations, with the younger one formed during the merger. In order to lose the
remaining gas, the dwarf galaxy must have additionally experienced some interaction with its host, Andromeda, whose hot
gaseous halo would strip the dwarf's gas via ram pressure. It is worth emphasizing that tidal stirring is unable to
induce significant stable rotation around the longest axis. Even if it occurs, it is always much smaller than the
velocity dispersion, while in Andromeda II these two quantities are comparable. We have also shown that prolate
rotation can be obtained from less idealized initial conditions, i.e. from dwarfs merging on non-radial orbits and with
different inclinations of the disks (\cite[Ebrov\'{a} \& {\L}okas 2015]{Ebrova2015}).

The scenario has been placed in the cosmological context although for more massive galaxies due to resolution
limitations. Almost 60 galaxies with prolate rotation have been identified in the Illustris simulation and most of them
were confirmed to be the result of a major merger (\cite[Ebrov\'{a} \& {\L}okas 2017]{Ebrova2017}). In most of them a
very tight correlation was found between the time of merger and the time of transition to prolate rotation and prolate
shape. In massive galaxies prolate rotation tends to occur only in gas-poor objects.

Although originally believed to be relatively simple, near-spherical, old-population, pressure-supported systems, dSph
galaxies of the Local Group only now start to reveal their complexity. Andromeda II has been recently shown to host
at least two stellar populations with different kinematics. While the older population shows prolate
rotation, the younger one rotates around the minor axis (\cite[del Pino et al. 2017b]{delPino2017b}). Even larger
degree of complexity has been discovered in the well-studied Fornax dwarf. The multiple stellar populations of Fornax
display different distributions and kinematics with the youngest stars centrally concentrated and showing a distinct
disky shape. Tracing the kinematics of different stellar populations as a function of age leads to the conclusion that
also Fornax probably experienced a merger about 8 Gyr ago (\cite[del Pino et al. 2017a]{delPino2017a}).

\section{Conclusions}

DSph galaxies of the Local Group remain an interesting subject of investigation. Determining the dark matter
distribution in dSph galaxies remains challenging but promising pathways for improvement exist. In the era of large
surveys we can expect the kinematic samples for the dwarfs to increase dramatically allowing for the modelling methods
to reach the next level of refinement. These should include in the near future non-spherical orbit-based modelling and
attempts to constrain the inner slope of the dark matter distribution which remained elusive so far. The measurements
of proper motions for a significant number of bright stars in dSphs should put additional constraints on the models,
both in terms of dark matter distribution and anisotropy of stellar orbits. These studies could also benefit from the
use of multiple stellar populations that at least in principle should be able to constrain dark matter distribution at
different scales.

Many dwarfs of the Local Group show signs of interactions that shaped them. The two dominant scenarios of such
interactions are the tidal stirring in the vicinity of the Milky Way and Andromeda, and mergers between dwarfs. For
example, tidal stirring can account for the non-spherical shape and remnant rotation in Carina and a merger can explain
the prolate rotation in Andromeda II. More advanced models created in close relation to recent observational findings
are needed to reproduce the properties of these quite complicated systems in more detail.

\section*{Acknowledgments}
This contribution was supported in part by the Polish National Science Center under grant 2013/10/A/ST9/00023.

\end{document}